\documentclass[twocolumn,english,aps,reprint,twocolumn,superscriptaddress]{revtex4-2}
\usepackage[T1]{fontenc}
\usepackage[latin9]{inputenc}
\usepackage{units}
\usepackage{amsmath}
\usepackage{amssymb}
\usepackage{graphicx}
\usepackage{esint}

\makeatletter
\usepackage[english]{babel}
\def\urlprefix{}
\def\url#1{}
\addto\captionsenglish{%
}

\makeatother

\usepackage{babel}
\begin{document}
\title{Discrete Quantum Geometry and Intrinsic Spin Hall Effect}
\author{Jie-Xiang Yu}
\affiliation{Department of Physics and Astronomy, University of New Hampshire,
Durham, New Hampshire 03824, USA}
\affiliation{Department of Physics, Center for Molecular Magnetic Quantum Materials
and Quantum Theory Project, University of Florida, Gainesville, Florida
32611, USA}
\author{Jiadong Zang}
\affiliation{Department of Physics and Astronomy, University of New Hampshire,
Durham, New Hampshire 03824, USA}
\author{Roger K. Lake}
\affiliation{Department of Electrical and Computer Engineering, University of California,
Riverside, California 92521, USA}
\author{Yi Zhang}
\affiliation{International Center for Quantum Materials, School of Physics, Peking
University, Beijing 100871, China}
\author{Gen Yin}
\thanks{gen.yin@georgetown.edu}
\affiliation{Department of Physics, Georgetown University, Washington, D.C. 20057,
USA}
\begin{abstract}
We show that the quantum geometry of the Fermi surface can be numerically
described by a 3-dimensional discrete quantum manifold. This approach
not only avoids singularities in the Fermi sea, but it also enables
the precise computation of the intrinsic Hall conductivity resolved
in spin, as well as any other local properties of the Fermi surface.
The method assures numerical accuracy when the Fermi level is arbitrarily
close to singularities, and it remains robust when Kramers degeneracy
is protected by symmetry. The approach is demonstrated by calculating
the anomalous Hall and spin Hall conductivities of a 2-band lattice
model of a Weyl semimetal and a full-band \emph{ab-initio} model of
zincblende GaAs. 
\end{abstract}
\maketitle
It has been well established that the intrinsic anomalous (spin) Hall
effect originates from the topological property of the Fermi sea,
and it can be evaluated through an integral of the Berry curvature
among the occupied states \citep{nagaosa_anomalous_2010,xiao_berry_2010}.
Such an approach has been employed for first-principles calculations
in different ways \citep{yao_first_2004,wang_ab_2006,guo_intrinsic_2008}.
A particularly convenient approach is Wannier interpolation \citep{marzari_maximally_1997},
where a dense mesh in $k$ space does not significantly increase the
numerical complexity. One difficulty of such an approach are the singularities
caused by band crossings. Such crossings, either accidental or protected
by symmetry, result in sharp peaks of the anomalous Hall conductivity.
This makes the numerical evaluation difficult, especially when the
singularities are close to the Fermi surface, which usually cannot
be avoided in a scan of the Fermi level.

\begin{figure}
\begin{centering}
\includegraphics[width=1\columnwidth]{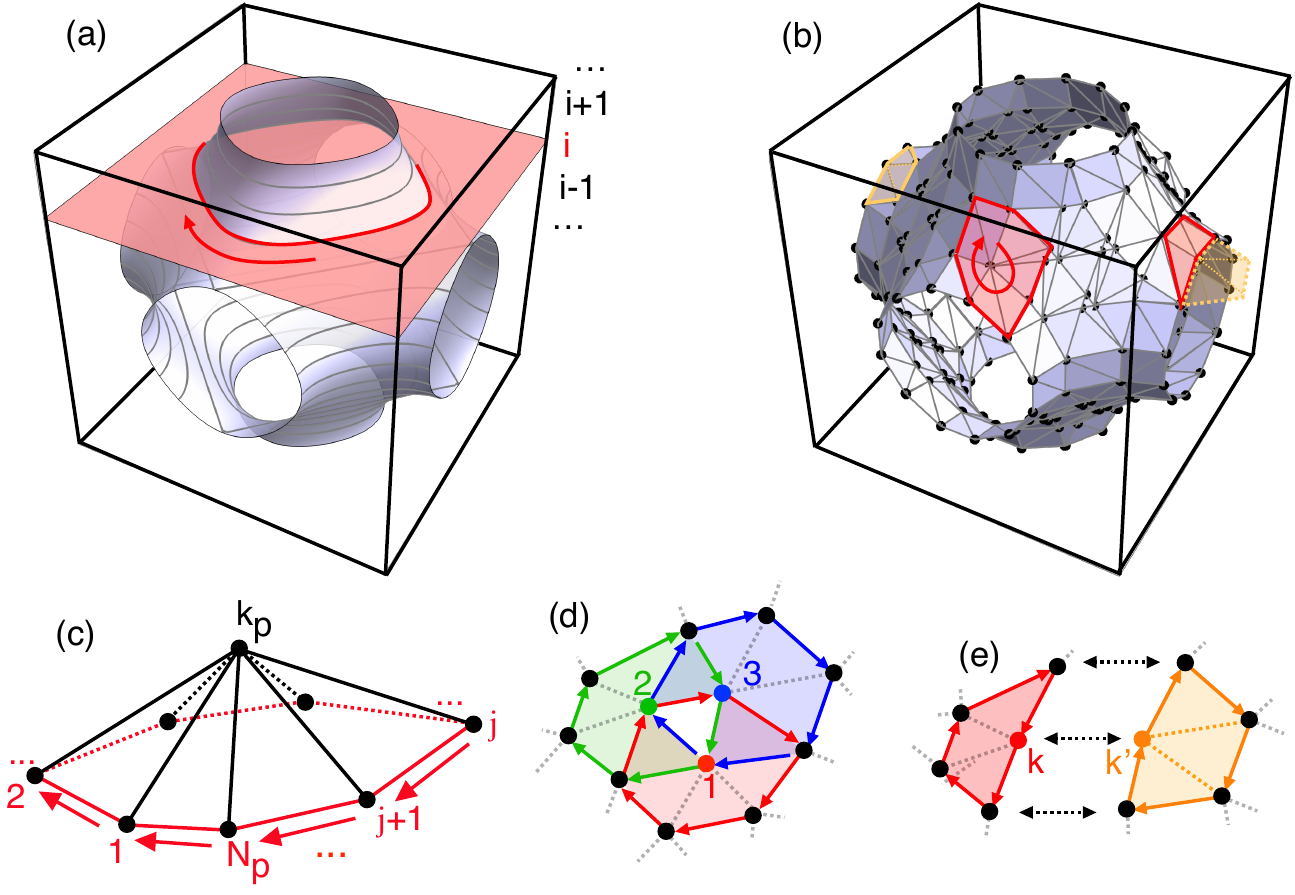} 
\par\end{centering}
\caption{ Illustration of different approaches to describe intrinsic anomalous
Hall effect as a geometric property of the Fermi surface. (a) The
slice-based approach. The Fermi surface ($\epsilon_{F}=0$) corresponds
to a cubic tight-binding model $\epsilon_{\mathbf{k}}=-2t(\cos k_{x}+\cos k_{y}+\cos k_{z})$,
where $t$ is the hopping between nearest neighbors. (b) The new vertex-based
method by establishing a three-dimensional discrete manifold of the
Fermi surface. (c) The local Berry phase associated with each vertex
given by the adjacent triangles. (d) The flux of Berry curvature is
over-counted by $3$ times for $\triangle_{123}$. (e) Manifold matching
between vertices and edges on opposing Brillouin-zone boundaries.
The red and orange polygons correspond to the ones shown in (b) near
the boundaries. \label{fig:Illustration-of-different-schemes}}
\end{figure}

An alternative and natural perspective is to describe the non-quantized
part of the anomalous Hall effect as a geometric property of the Fermi
surface \citep{haldane_berry_2004}. This approach formulates the
anomalous Hall conductivity as a surface integral of Berry curvature
weighted by the Fermi wave vector $\mathbf{k}_{F}$, which naturally
avoids singularities in the Fermi sea. The method has been implemented
for full-band models by decomposing the integral on the Fermi surface
into a sum of oriented Wilson loops \citep{wang_fermi-surface_2007,gosalbez-martinez_chiral_2015}:
$\sigma_{xy}\sim\sum_{n,i}k_{z,i}\phi_{n,i}$. Here, $n$ is the band
index, $k_{z,i}$ is the $z$ component of $\mathbf{k}_{F}$ on the
$i\textrm{-th}$ slice, and $\phi_{n,i}=\ointclockwiseop^{i}\mathbf{A}_{n}\cdot d\mathbf{l}$
is the gauge-invariant Berry phase looping through the $i\textrm{-th}$
slice as schematically shown in Fig. \ref{fig:Illustration-of-different-schemes}(a).
At the continuous limit, $\mathbf{A}_{n}=i\left\langle \psi_{n}^{\mathbf{k}}\right|\nabla_{\mathbf{k}}\left|\psi_{n}^{\mathbf{k}}\right\rangle $
is the Berry connection of the $n\textrm{-th}$ band. Numerically,
the closed loops are decomposed into discrete Wilson links, such that
$\phi_{n,i}=-\textrm{Im}[\ln\prod_{j}\langle j|j+1\rangle]$, where
$\langle j|j+1\rangle$ is the inner product between two neighboring
eigenstates associated with a segment of the loop. Although $\phi_{n,i}$
is defined on $(-\pi,\pi]$, which seems arbitrary, one can enforce
a smooth $\phi_{n,i}$ by carefully handling the branches of the Berry
phase, such that $|\phi_{n,i}-\phi_{n,i+1}|\ll2\pi$. Thus, the Fermi-surface
approach only contains an ambiguity with multiples of $2\pi$ compared
to the Fermi-sea prescription, which corresponds to the undetermined
quantized part of the Hall conductivity. 

Although the aforementioned slice-based approach is promising, it
cannot capture the intrinsic spin Hall effect. Obtained by the loop
integral on each slice of the Fermi surface, the Berry phase is formulated
as a global property. Namely, the spin-Hall conductivity can only
be expressed as $\sigma_{xy}^{(\alpha)}\sim\sum_{n,i}s_{\alpha,i}k_{z,i}\phi_{n,i}$
where $s_{\alpha,i}$ is the uniform spin on slice $i$. This becomes
problematic when strong spin-orbit coupling (SOC) occurs, where spins
usually have a texture on the Fermi surface, and therefore must be
evaluated locally. Furthermore, the slice-based approach requires
the choice of $k_{z}$ perpendicular to the $x\textrm{-}y$ transport
plane. Any rotation of the transport plane requires re-establishing
the slices and re-evaluating the Berry phases $\{\phi_{i}\}$ for
all slices. This is counterintuitive since the Fermi surface should
contain enough information to determine the intrinsic (spin) Hall
conductivity for arbitrary current directions.

In this paper, we show that the non-quantized part of the intrinsic
anomalous (spin) Hall conductivity can be fully captured by a three-dimensional
(3D) discrete quantum manifold of the Fermi surface. In such a manifold,
Wilson links form local loops, as shown in Fig. \ref{fig:Illustration-of-different-schemes}(b),
and therefore the Hall conductivity can be resolved in spin, as well
as any other local properties of the Fermi surface. Such an approach
can naturally handle singularities near the Fermi surface, and it
remains robust when Kramers degeneracy is protected by symmetry. The
approach is tested on a $2\times2$ tight-binding model of Weyl semimetal
and an \emph{ab-initio} full-band model of zincblende GaAs.

At the limit of linear response, the nonequilibrium group velocity
driven by an applied electric field $\boldsymbol{{\cal E}}$ contains
two parts: $\mathbf{v}=\mathbf{v}_{0}+\mathbf{v}_{a}$, where $\mathbf{v}_{0}=\frac{1}{\hbar}\nabla_{\mathbf{k}}\epsilon$
is the regular group velocity and $\mathbf{v}_{a}=\frac{e}{\hbar}\boldsymbol{{\cal E}}\times\boldsymbol{\Omega}_{n}\left(\mathbf{k}\right)$
is the anomalous velocity. At zero temperature, the leading-order
average of $\mathbf{v}_{a}$ is 
\begin{equation}
\langle\mathbf{v}_{a}\rangle=\frac{e\boldsymbol{{\cal E}}}{\rho\hbar}\times\left(\frac{1}{2\pi}\right)^{3}\sum_{n}\boldsymbol{I}_{n}\label{eq:AverageAnomalousVelocity}
\end{equation}
where $\rho$ is the carrier density and $n$ is the band index. Thus,
\begin{equation}
\boldsymbol{I}_{n}=\int_{\textrm{BZ}}f_{0}\boldsymbol{\Omega}_{n}dv=\int_{\textrm{BZ}}f_{0}\left(\nabla\times\mathbf{A}_{n}\right)dv,\label{eq:UnitlessIntegral}
\end{equation}
where $f_{0}$ is the equilibrium Fermi factor. At zero temperature,
using integration by parts, 
\begin{equation}
\boldsymbol{I}_{n}=\sum_{\alpha}\hat{n}_{\alpha}\times\int_{\textrm{BZB}}\left(f_{0}\mathbf{A}_{n}\right)ds_{\alpha}+\int_{\textrm{BZ}}\delta\left(\epsilon-\epsilon_{F}\right)\boldsymbol{\nabla}\epsilon\times\mathbf{A}_{n}dv\label{eq:HaldaneIntegral}
\end{equation}
where the first integral goes over Brillouin-zone boundaries (BZBs),
and the second one is a volume integral within the Brillouin zone
(BZ). Eq. (\ref{eq:HaldaneIntegral}) is equivalent to the one by
Haldane \citep{haldane_berry_2004}. Here, $\hat{n}_{\alpha}$ ($\alpha=1,2,\cdots,6$)
is the normal vector of BZBs. For 3D crystals, one can choose a Brillouin
zone with three pairs of opposing BZBs. For each pair, $\hat{n}_{\alpha}=-\hat{n}_{\beta}$,
and the first term in Eq. (\ref{eq:HaldaneIntegral}) vanishes. As
a result, 
\begin{equation}
\boldsymbol{I}_{n}=\int_{\textrm{FS}}d\mathbf{s}\times\mathbf{A}_{n}.\label{eq:SurfaceIntegral}
\end{equation}
The manifold defined on a continuous Fermi surface is topologically
equivalent to a discrete manifold as shown in Fig. \ref{fig:Illustration-of-different-schemes}(b).
Such a manifold can be obtained by the marching tetrahedra method.
Once an intersection is detected between $\epsilon_{F}$ and a tetrahedron,
the Fermi-surface piece must be an oriented triangle, since the normal
vector can be defined along $\nabla_{\mathbf{k}}\epsilon_{F}$. The
orientation of each triangle distinguishes electron and hole bands.
The integration in Eq. (\ref{eq:SurfaceIntegral}) can thus be approximated
by summing through the surface pieces: 
\begin{equation}
\overrightarrow{\triangle}=\frac{1}{2}\left(\mathbf{k}_{2}-\mathbf{k}_{1}\right)\times\left(\mathbf{k}_{3}-\mathbf{k}_{1}\right)\times\mathbf{A}\approx\hat{n}\times\mathbf{A}ds\label{eq:TriangleComponent}
\end{equation}
where $\hat{n}$ is the normal vector, and the three vertices of the
triangle are labeled by $1,2,3$, following some convention of chirality
with respect to $\hat{n}$. With some algebra, 
\begin{align}
\overrightarrow{\triangle} & =\frac{1}{2}\left(w_{12}\mathbf{k}_{3}+w_{23}\mathbf{k}_{1}+w_{31}\mathbf{k}_{2}\right)\label{eq:TriangleWilsonLinks}\\
 & =\frac{1}{2}\mathbf{k}_{1}\left(w_{12}+w_{23}+w_{31}\right)+o\left(\mathbf{k}_{1}\right).
\end{align}
Here,
\begin{equation}
w_{\alpha\beta}=-\textrm{Im}\left[\mathbf{A}\cdot\left(\mathbf{k}_{\beta}-\mathbf{k}_{\alpha}\right)\right]=-\textrm{Im}\left(\ln\langle\mathbf{k}_{\alpha}|\mathbf{k}_{\beta}\rangle\right)\label{eq:EachWilsonLink}
\end{equation}
is the Wilson link between two adjacent vertices, where $|\mathbf{k}_{\alpha}\rangle$
and $|\mathbf{k}_{\beta}\rangle$ are the corresponding eigenstates.
The integral in Eq. (\ref{eq:SurfaceIntegral}) can therefore be written
as 
\begin{equation}
\boldsymbol{I}_{n}=\sum_{i}\overrightarrow{\triangle}_{n,i}=\frac{1}{6}\sum_{p}\mathbf{k}_{n,p}\phi_{n,p}\label{eq:I_n}
\end{equation}
where $i$ labels triangles and $\mathbf{k}_{n,p}$ is the k-space
position of vertex $p$ on in the $n\textrm{-th}$ shell of the Fermi
surface. Here, $\phi_{n,p}$ is the gauge-invariant local Berry phase
\begin{equation}
\phi_{n,p}=-\textrm{Im}\left(\ln\prod_{j}^{N_{p}}\langle j,n|j+1,n\rangle\right),\label{eq:I_n_Log}
\end{equation}
where $j$ goes through the vertices surrounding vertex $p$ as shown
in Fig. \ref{fig:Illustration-of-different-schemes}(c). The extra
factor of $1/3$ in Eq. (\ref{eq:I_n}) comes from over-counting the
Berry phase as shown in Fig. \ref{fig:Illustration-of-different-schemes}(d).
The logarithm in Eq. (\ref{eq:I_n_Log}) is defined on the branch
of $(-\pi,\pi]$ with an unknown multiple of $2\pi$. Such ambiguity
can be naturally ruled out since the local Berry phase is close to
zero piecewise. The Hall conductivity can then be obtained by $\sigma_{\mu\nu}=\frac{e^{2}}{\hbar}\left(\frac{1}{2\pi}\right)^{3}\sum_{n}\boldsymbol{I}_{n}\cdot\hat{\lambda}$
for an arbitrary coordinate setup, where $\mu$, $\nu$ and $\lambda$
represent $x$, $y$ and $z$ in cyclic order.

The summation in Eq. (\ref{eq:I_n}) holds true for both the vertices
inside the Brillouin zone and the ones on the BZBs. Defined in the
affine $\mathbf{k}$ space, the Fermi surface is periodic and the
surface edges on opposing BZBs are therefore connected. This requires
careful handling to ensure translational symmetry. As shown in Fig.
\ref{fig:Illustration-of-different-schemes}(e), each boundary vertex
$\mathbf{k}$ must paired with another vertex $\mathbf{k}'$ on the
opposing BZB, shifted by a reciprocal lattice vector. As a result,
a closed Wilson loop can be formed for each boundary vertex by including
the incoming and outgoing edges on the BZB. Mismatching of boundary
vertices will result in topological defects of the discrete manifold,
making it non-homeomorphic with the continuous limit. Similar to the
slice-based approach, Eq. (\ref{eq:I_n}) also leaves the quantized
part of Hall conductivity undetermined. This can be seen by taking
$\mathbf{k}\rightarrow\mathbf{k}+\mathbf{G}$ for all vertices, where
$\mathbf{G}$ is an arbitrary reciprocal lattice point. 

The approach defined by Eqs. (\ref{eq:I_n}-\ref{eq:I_n_Log}) enables
the evaluation of intrinsic spin-Hall effect for materials with strong
SOC. Since Eq. (\ref{eq:I_n}) sums through all vertices, each vertex
can be resolved in any local properties such as the electron spin.
As a result the intrinsic spin Hall conductivity can be straightforwardly
obtained using 
\begin{equation}
\left\langle \mathbf{v}_{a}s_{\alpha}\right\rangle =\frac{e\boldsymbol{{\cal E}}}{\rho\hbar}\times\left(\frac{1}{2\pi}\right)^{3}\sum_{n}\boldsymbol{I}_{n}^{(\alpha)}.\label{eq:SpinVelocity}
\end{equation}
Here, $\left\langle \mathbf{v}_{a}s_{\alpha}\right\rangle $ is the
expectation of the spin-velocity tensor, where $s_{\alpha}$ represents
the Pauli matrices for electron spins ($\alpha=x,y,z$). For each
band, the integral $\boldsymbol{I}{}_{n}^{(\alpha)}$ can be obtained
by summing through all vertices $\boldsymbol{I}_{n}^{(\alpha)}=\frac{1}{6}\sum_{p}\mathbf{k}_{n,p}\phi_{n,p}\langle s_{\alpha,p}\rangle$,
where $n$ is the band index and $\langle s_{\alpha,p}\rangle$ is
the spin expectation on vertex $p$. The spin Hall conductivity is
then written as $\sigma_{\mu\nu}^{(\alpha)}=\left(\frac{1}{2\pi}\right)^{3}\frac{e}{2}\sum_{n}\boldsymbol{I}_{n}^{(\alpha)}\cdot\hat{\lambda}$.

\begin{figure}[t]
\begin{centering}
\includegraphics[width=1\columnwidth]{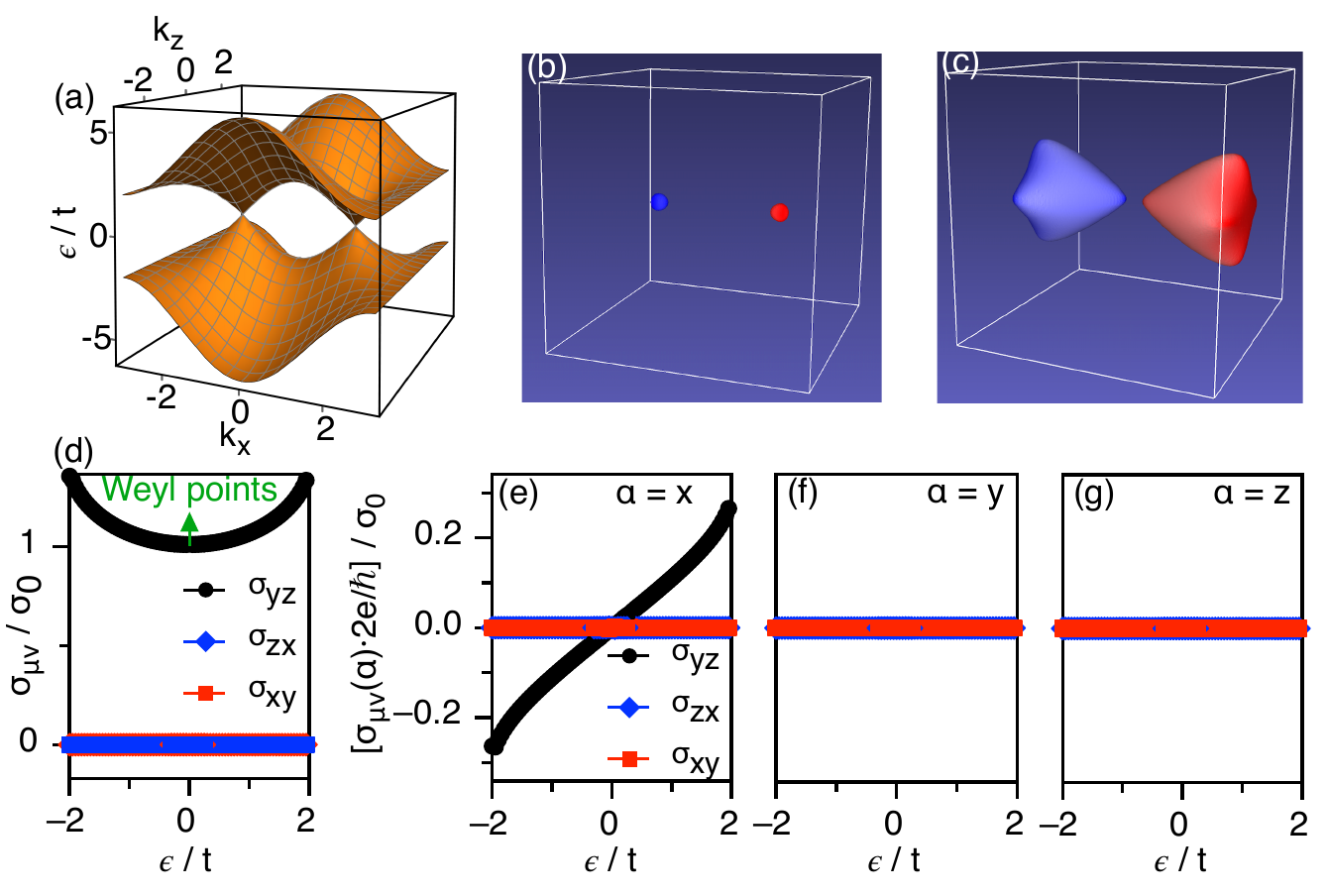} 
\par\end{centering}
\caption{Intrinsic (spin) Hall effect in a lattice model of a Weyl semimetal.
(a) The band structure given by the model with $k_{y}=0$, $k_{0}=\frac{\pi}{2}$,
$t_{x}=t$, $m=2t$. (b) Fermi surface obtained as a 3D discrete quantum
manifold ($\epsilon_{F}=0.45t$). The vertices are colored by the
local Berry curvature. (c) The Fermi surface at $\epsilon_{F}=1.05t$,
where the shape represents the symmetry of the lattice. (d) The intrinsic
anomalous Hall conductivity computed using Eqs. (\ref{eq:I_n}) -
(\ref{eq:I_n_Log}). Here, $\sigma_{\mu\nu}$ is normalized to the
Weyl-point conductivity $\sigma_{0}=\nicefrac{e^{2}}{2h}$. (e-f)
The intrinsic spin-Hall conductivity tensor evaluated using Eq. (\ref{eq:SpinVelocity}).
The results are first converted to the corresponding electrical conductivity
and then normalized to $\sigma_{0}$.}
\label{fig:Proof-of-concept} 
\end{figure}

The vertex-based Fermi-surface approach remains well-defined when
$\epsilon_{F}$ is arbitrarily close to point-singularities of Berry
curvature, i.e. Weyl points. We demonstrate this using a $2\times2$
tight-binding model of a Weyl semimetal: 
\begin{align}
H & =s_{x}\left[m\left(2-\cos k_{y}-\cos k_{z}\right)+2t_{x}\left(\cos k_{x}-\cos k_{0}\right)\right]\nonumber \\
 & -2ts_{y}\sin k_{y}-2ts_{z}\sin k_{z}.\label{eq:Weyl2x2Lattice}
\end{align}
The first term breaks time-reversal symmetry, resulting in two Weyl
points at $(\pm k_{0},0,0)^{T}$ as shown in Fig. \ref{fig:Proof-of-concept}(a).
When $\epsilon_{F}$ is close to the Weyl points, the Fermi surface
contains approximately two spheres, each wrapping a sink and a source
of the Berry curvature, respectively {[}Fig. \ref{fig:Proof-of-concept}(b){]}.
Away from the Weyl points, the Fermi surface is shown in Fig. \ref{fig:Proof-of-concept}(c),
where the symmetry of the cubic lattice can be seen. Computed from
Eq. (\ref{eq:I_n}), anomalous Hall conductivities along different
directions are shown in Fig. \ref{fig:Proof-of-concept}(d). As expected
\citep{yang_quantum_2011}, the Hall conductivity is $\sigma_{0}=\frac{e^{2}}{\pi h}k_{0}=\frac{e^{2}}{2h}$
when $\epsilon_{F}$ is aligned with the Weyl points. From the perspective
of the discrete manifold, this can be understood by noticing $\mathbf{k}_{n,p}\approx(\pm\frac{\pi}{2},0,0)^{T}$,
which can be factored out when $\epsilon_{F}\rightarrow0$. The summation
in Eq. \ref{eq:I_n} then becomes the total flux of the Berry curvature
evaluated on two spheres ($\phi_{\pm}=\pm2\pi$) weighted by constants
$\pm k_{0}$, respectively. Numerically, the spheres are resolved
as two tetrahedra at the limit of $\epsilon_{F}\rightarrow0$. However,
since these tetrahedra remain homeomorphic with spheres, the numerical
evaluation of the Berry-curvature flux is only determined by the number
of sources or drains enclosed. This allows for precise numerical results
when $\epsilon_{F}$ is arbitrarily close to the Weyl points. Using
Eq. \ref{eq:SpinVelocity}, one can obtain all the elements of the
spin Hall conductivity tensor $\sigma_{\mu\nu}^{(\alpha)}$, as shown
in Figs. \ref{fig:Proof-of-concept}(e-g). This suggests that $s_{x}$
has a spin Hall effect when transporting in the $y\textrm{-}z$ plane.
Such spin Hall effect is only allowed when $\epsilon_{F}$ is shifted
away from the Weyl points.

The idea of a discrete quantum manifold can be directly applied for
\emph{ab-initio} models. Here we demonstrate such application on the
full-band model of zincblende GaAs. The model is obtained using a
spin-polarized first-principles calculation with the projector augmented
wave pseudopotential \citep{blochl_projector_1994} implemented in
Vienna Ab Initio Simulation Package (VASP) \citep{kresse_efficiency_1996,kresse_efficient_1996}.
The primitive unit cell of the face-centered cubic lattice contains
one Ga atom and one As atom. The corresponding Brillouin zone is sampled
with a $\Gamma$-centered $7\times7\times7$ grid and an energy cutoff
of $500\,\textrm{eV}$ is applied for the plane-wave expansion. The
generalized gradient approximation (GGA) is applied using the Perdew-Burke-Ernzerhof
(PBE) form \citep{perdew_generalized_1996} as the exchange-correlation
functional to obtain the Kohn-Sham eigenstates. A partially self-consistent
$\mathrm{GW}$ approximation \citep{hybertsen_electron_1986,shishkin_implementation_2006,shishkin_self-consistent_2007}
is employed to obtain the quasi-particle energies from the fixed eigenstates.
Spin-orbit coupling is included. To construct the tight-binding Hamiltonian
in a Wannier-function (WF) basis, we performed a unitary transformation
\citep{marzari_maximally_1997} implemented in Wannier90 \citep{mostofi_updated_2014}.
Time-reversal and other symmetries of the tight-binding model are
enforced by using the package TB-models \citep{gresch_automated_2018}.
The tight-binding Hamiltonian has the same exact energies as those
obtained from VASP in the energy window from the lowest eigenvalues
to $3.5\thinspace\textrm{eV}$ above the Fermi energy. 
\begin{figure}
\begin{centering}
\includegraphics[width=1\columnwidth]{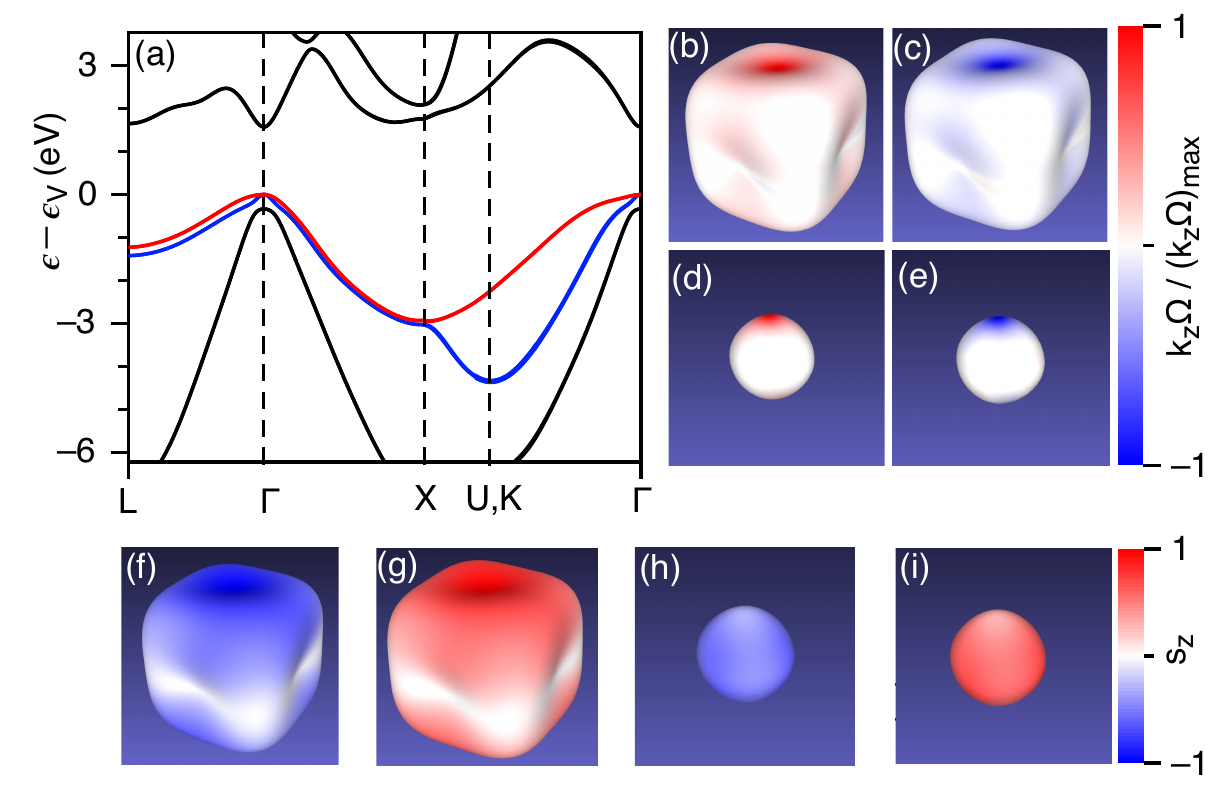} 
\par\end{centering}
\caption{(a) The band structure of GaAs along high-symmetry points. Both the
heavy hole (red) and the light hole (blue) bands are 2-fold degenerate.
(b-e) The 3D discrete quantum manifold of the Fermi surface at $\epsilon=\epsilon_{V}-0.02\thinspace\textrm{eV}$.
(b) and (c) correspond to the heavy hole band, whereas (d) and (e)
correspond to the light hole band. (f-i) The electron spin $s_{z}$
illustrated on the Fermi surfaces corresponding to (b-e), respectively.
\label{fig:GaAs}}
\end{figure}

\begin{figure}
\begin{centering}
\includegraphics[width=1\columnwidth]{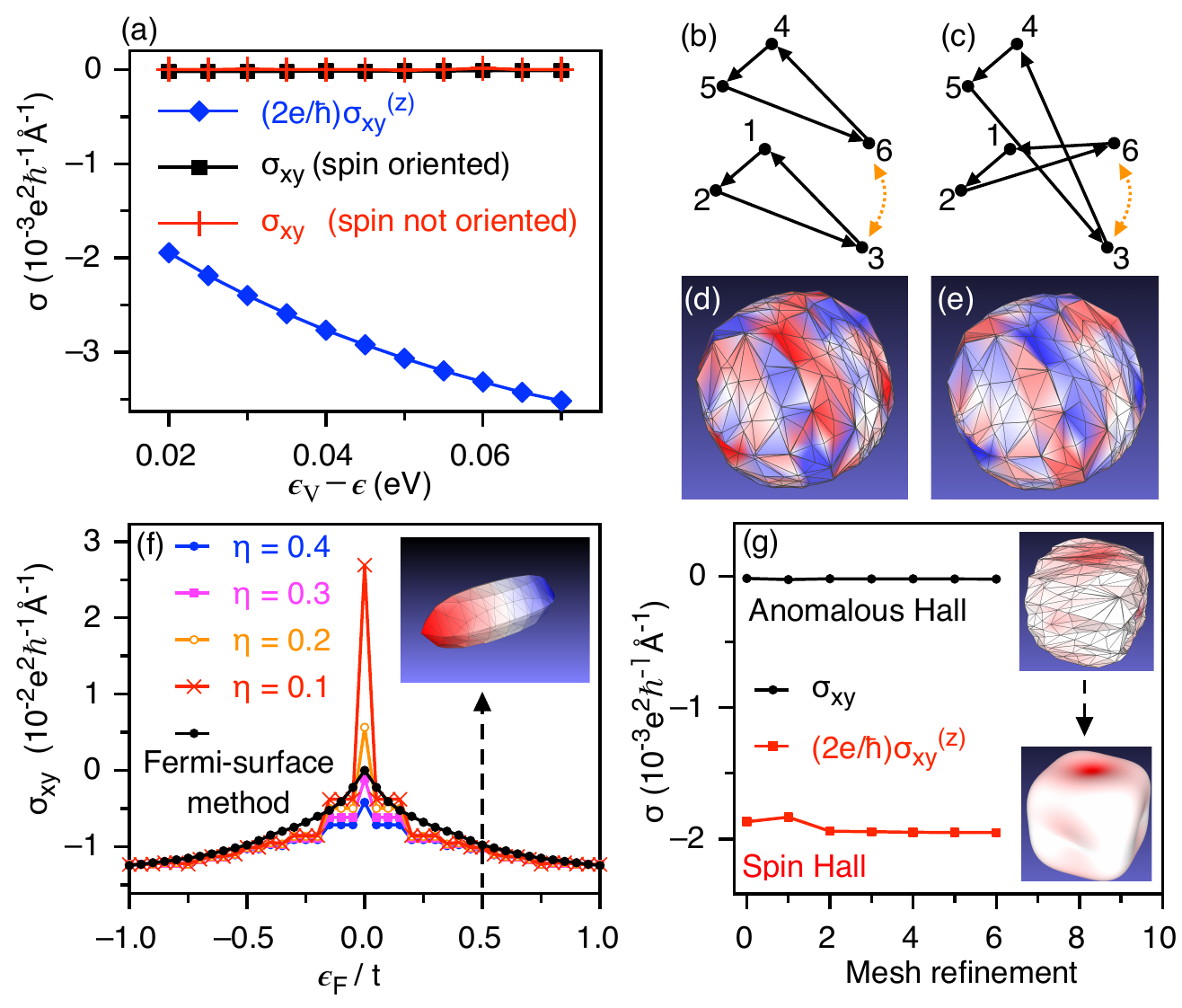} 
\par\end{centering}
\caption{(a) The anomalous Hall conductivity (dark) and the intrinsic spin
Hall conductivity (blue) near the valence band top of zincblende GaAs.
The red line denotes the Hall conductivity without orienting the spins
of degenerate states. (b) Two coincident triangles due to Kramers
degeneracy. The illustration is shifted for demonstration. (c) Swapping
one Kramers pair in the triangle. (d-e) The discrete heavy-hole Fermi
surface at $\epsilon_{F}=\epsilon_{V}-0.02\thinspace\textrm{eV}$
without orienting the degenerate spins. The vertex color illustrates
the local Berry phases $\phi_{p}$. (f) Comparison between the Fermi-sea
and the Fermi-surface results. The inset shows the Fermi-surface at
$\epsilon_{F}=0.5t$. (g) The convergence of numerical results after
mesh refinements. The Insets compare the $0\textrm{-th}$ and the
$6\textrm{-th}$ refinement for the heavy-hole band of GaAs. \label{fig:Intrinsic-(spin)-Hall}}
\end{figure}

The band structure of zincblende GaAs is shown in Fig. \ref{fig:GaAs}(a),
where the valence band splits into heavy-hole, light-hole and split-off
bands due to SOC. Each band contains a 2-fold Kramers degeneracy.
The Fermi surfaces at $\epsilon_{F}=\epsilon_{V}-0.02\thinspace\textrm{eV}$
are shown in Figs. \ref{fig:GaAs}(b-e), with the vertices colored
by the $k_{z}$-weighted Berry curvature evaluated as $\phi_{p}k_{p,z}/\delta s_{p}\approx\left(\text{\ensuremath{\Omega}}\cdot d\hat{s}\right)k_{z}$.
Here, $\delta s_{p}$ is the total area of the triangles surrounding
vertex $p$, whereas $d\hat{s}$ is the normal direction of the Fermi
surface at the continuous limit. As expected, the Hall conductivities
given by the Kramers pairs have opposite signs and exactly cancel
each other. However, since the Kramers pairs have opposite spins as
shown in Figs. \ref{fig:GaAs}(f-i), the spin Hall conductivity should
be finite. This is consistent with previous results \cite{guo_ab_2005,qi_topological_2006,murakami_dissipationless_2003}.
Here, the spins of the Kramers pairs are re-oriented by diagonalizing
$\gamma_{z}=\mathbf{I}\otimes s_{z}$ within the degenerate subspace.
The updated eigenstates are then sorted based on the eigenvalues of
$\gamma_{z}$ \cite{solovyev_validity_2015}. Such operation is necessary
because arbitrary superpositions within Kramers degeneracy result
in arbitrary spin directions, and the spin Hall effect is thus ill-defined.
In experiments, the choice of the spin direction is determined by
the measurement setup. With the spins properly oriented for the Kramers
pairs, the (spin) Hall conductivities at different positions of $\epsilon_{F}$
are illustrated in Fig. \ref{fig:Intrinsic-(spin)-Hall}(a). As $\epsilon_{F}$
shifts deeper into the valence band, $k_{F}$ increases for all Fermi
surfaces, therefore the magnitude of $\sigma_{xy}^{(z)}$ also increases. 

Although spins are required to properly orient for the spin Hall effect,
such a requirement is not necessary for the anomalous Hall effect.
Without diagonalizing the spin, the degenerate eigenstates are randomly
ordered. Here we show that the order does not affect the anomalous
Hall conductivity evaluated by the discrete quantum manifold. Due
to Kramers degeneracy, each piece of the Fermi surface contains two
coincident triangles in $k$ space. We label these triangles $\triangle_{123}$
and $\triangle_{456}$ as shown in Fig. \ref{fig:Intrinsic-(spin)-Hall}(b).
Here, ($|1\rangle$,$|4\rangle$), ($|2\rangle$,$|5\rangle$) and
($|3\rangle$,$|6\rangle$) correspond to three Kramers pairs. These
pairs must be protected by an antiunitary transformation ${\cal K}$,
such that $|1\rangle=|{\cal K}4\rangle$, $|2\rangle=|{\cal K}5\rangle$
and $|3\rangle=|{\cal K}6\rangle$. As a result, $\langle1|2\rangle=\langle{\cal K}4|{\cal K}5\rangle=\langle4|5\rangle^{*}$,
hence $w_{12}=-w_{45}$. Such relation holds for all coinciding edges
of the two triangles, and therefore $\phi_{123}=w_{12}+w_{23}+w_{31}=-(w_{45}+w_{56}+w_{64})=-\phi_{456}$.
Without losing generality, we swap $|3\rangle$ and $|6\rangle$,
such that the two triangles become $\triangle_{126}$ and $\triangle_{345}$,
as illustrated in Fig. \ref{fig:Intrinsic-(spin)-Hall}(c). Thus,
$\langle2|6\rangle=\langle{\cal K}5|6\rangle=\langle5|{\cal K}6\rangle^{*}=\langle5|3\rangle^{*}$,
such that $w_{26}=-w_{53}$. Similarly, $w_{61}=-w_{34}$, and hence
$\phi_{126}=-\phi_{345}$. Note that in general $\phi_{123}\ne\phi_{126}$
and $\phi_{456}\ne\phi_{345}$. However, since $\phi_{123}+\phi_{456}=\phi_{126}+\phi_{345}=0$,
the anomalous Hall conductivity is strictly suppressed. This can be
seen from Figs. \ref{fig:Intrinsic-(spin)-Hall}(d-e) where the local
Berry phases are illustrated by the vertex colors. Although the Berry
phases for each branch of the degenerate bands are assigned randomly,
their contributions to the anomalous Hall conductivity add to zero,
as shown by the red line in Fig. \ref{fig:Intrinsic-(spin)-Hall}(a). 

The Fermi-surface approach has better numerical accuracy compared
to the Fermi-sea approach, especially when $\epsilon_{F}$ is close
to the singularities. We demonstrate this by setting $k_{0}=0$ in
Eq. \ref{eq:Weyl2x2Lattice} such that the two Weyl points merge at
$\Gamma$ point, and the Hall conductivity should be zero when $\epsilon_{F}=0$.
For simplicity the $k$ values are based on the lattice constant of
$1\text{�}$. Using the Fermi-sea approach, the anomalous Hall conductivity
is given by $\sigma_{yz}=\frac{e^{2}}{\hbar}\left(\frac{1}{2\pi}\right)^{3}\int_{\epsilon<\epsilon_{F}}\Omega_{x,n}dv$,
where $\Omega_{x,n}=-2\textrm{Im}\sum_{m\neq n}\frac{\langle n_{\mathbf{k}}|\partial_{k_{y}}H|m_{\mathbf{k}}\rangle\langle m_{\mathbf{k}}|\partial_{k_{z}}H|n_{\mathbf{k}}\rangle}{(\epsilon_{n,\mathbf{k}}-\epsilon_{m,\mathbf{k}})^{2}+\eta^{2}}$.
Here, a finite $\eta$ has to be included in the denominator to avoid
singularity. Using a $32\times32\times32$ mesh, the Fermi-sea results
are compared to the Fermi-surface one in Fig. \ref{fig:Intrinsic-(spin)-Hall}(f).
When $\epsilon_{F}$ is away from the merged Weyl point, both approaches
obtain similar results. However, $\sigma_{yz}$ encounters a singularity
at $\epsilon_{F}=0$, and varies significantly with $\eta$. This
is not an issue for the Fermi-surface approach, where $\sigma_{yz}$
smoothly reduces to $0$. Another advantage of the Fermi-surface approach
is a better scalability. Since it only involves vertices near $\epsilon_{F}$,
the time complexity scales as $O(n^{2})$, where $n$ is the linear
mesh density. On the contrary, the Fermi-sea approach scales as $O(n^{3})$,
since the mesh needs to resolve a $3\textrm{D}$ volume. To demonstrate
the numerical stability of the Fermi-surface approach we compare the
(spin) Hall conductivities of GaAs at different levels of surface-mesh
refinement {[}Fig. \ref{fig:Intrinsic-(spin)-Hall}(g){]}. For simplicity
the Fermi surface is uniformly refined, where the number of triangles
doubles after each refinement.

The method introduced in this paper has been focusing on bulk materials
only. In principle, such approach is also suitable for 2D materials,
since one can introduce a third parameter $\lambda_{z}$, such that
$H(k_{x},k_{y},\lambda_{z})$ is periodic in all three dimensions
\citep{ryu_topological_2010,turner_quantized_2012}. The intrinsic
(spin) Hall conductivity will then be obtained `per thickness', treating
the new parameter as the $k_{z}$-dependency similar to the case of
a bulk. Although the Fermi surfaces demonstrated in this letter are
only resolved in spin, the method can, in principle, resolve the Hall
conductivity in any local properties of the Fermi surface such as
the angular momentum. New types of Hall effects can thus be straightforwardly
evaluated, suggesting the rich physics to further explore.

\textbf{Acknowledgements:} Work at UNH was supported by U.S. Department
of Energy, Office of Science, Basic Energy Sciences under No. DE-SC0020221.
First-principles calculations were conducted on Extreme Science and
Engineering Discovery Environment (XSEDE) under Grant No. TGPHY170023.

\end{document}